# ANOMALOUS PHASE DIAGRAMS IN THE SIMPLEST PLASMA MODELS

Igor L. Iosilevski, Alexander Yu. Chigvintsev

*Moscow Institute of Physics and Technology, (State University) Dolgoprudny 141700, Russia*

## INTRODUCTION

Problem of Phase Transition (PT) is of traditional great interest in astrophysics [1,2,3] as well as in general theory of so-called Strongly Coupled Coulomb Systems (SCCS) during very long time. Problem of Wigner crystallization in mixture of C(6+) and O(8+) nuclei during cooling of white dwarfs [2] there exists very interesting problem of stratified layers of high-Z crystals in outer core of neutron stars [4]. Besides crystallization the so-called hypothetical 'Plasma Phase Transition' (PPT) [5] and He/H2 phase decomposition [6] are examples of fluid-fluid phase transitions, which are of great interest for the theory of cooling of interiors of Giant Planets (GP) and Brawn Dwarfs (BD) [7] (see also paper [8]).

Besides the study of hypothetical PT in real plasmas a complementary approach is developing where the main subject of interest is definitely existing PT-s in simplified plasma models [9,10]. The well-known prototype model is OCP with a rigid background (notation – OCP(#)). This variant of OCP is studied carefully nowadays. The system cannot collapse or explode spontaneously. The only phase transition – crystallization – occurs in OCP(#) without any density change. More realistic model is One Component Plasma on uniform, but compressible compensating background (following notation – OCP(~)). One of the simplest example of OCP(~) is the model of classical point charges on uniform compressible background if ideal fermi-gas of electrons.

Transition to the OCP on uniform and compressible background leads to appearance of a new first-order phase transitions of gas-liquid type [10,11]. New phase diagram combines previous crystallization, now with a finite density change, with a qualitatively different coexistence curve of the new phase transition [12]. Obviously the structure and parameters of this phase diagram strongly depend on the value of charge number *Z*. This fact is illustrated at general phase diagram Figure 1.

# TOPOLOGY OF PHASE DIAGRAMS IN OCP(~) MODEL

Four qualitatively different situations should be distinguished for the OCP(~) depending on the value of charge number $Z$:

1) Low value of charge number $\qquad Z < Z_1^* \approx 35$
2) High value of charge number $\qquad Z > Z_2^* \approx 45$
3) Intermediate value of charge number $\qquad Z_1^* < Z < Z_2^*$
4) Boundary value of intermediate charge number interval $\qquad Z=Z_1^*$ and $Z= Z_2^*$

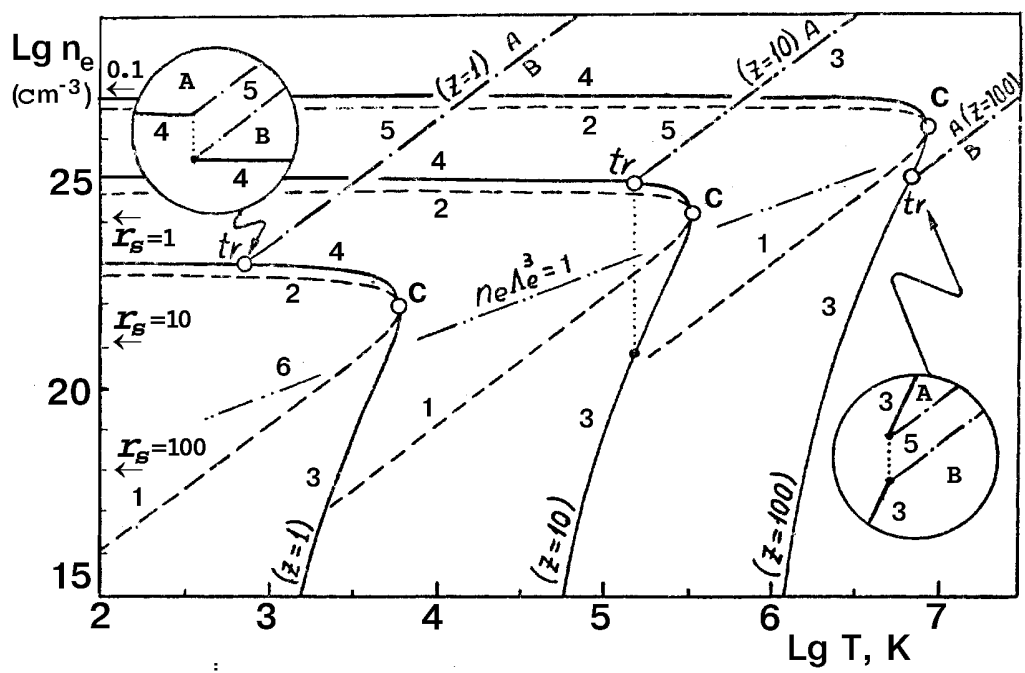

**FIGURE 1.** Phase diagram of OCP classical point ions on the compensating background of ideal fermi-gas of electrons in $T$–$n_e$ plane (temperature - background electron density) for $Z = 1$, 10 and 100. <u>Notations</u>: Spinodals (*1,2*) и binodals (*3,4*) of two-phase coexistence curve of condensed (*2,4*) and gaseous (*1,3*) phases; 5 – melting $\{\Gamma \equiv (Ze)^2/kTa \approx 175\}$, (*A* – crystal, *B* – fluid); Critical (*C*), triple (*tr*) points; 6 – boundary of electron degeneracy. Position of constant electron Brueckner parameter $r_S = 0.1$, 1, 10, and 100 are shown ($r_S = 100$ corresponds to the cold melting of electron Wigner crystal [13]).

## Low Values of Charge Number ($Z \sim 1$)

Phase diagram of the model was carefully studied in [10,11,12]. The *ordinary* structure of global phase diagram was obtained in this case: i.e. the relative position of critical and triple points, of melting zone and gas-liquid and gas-crystal coexistence, all are totally equivalent to those for ordinary substances (see Figure 2 and Figure 3).

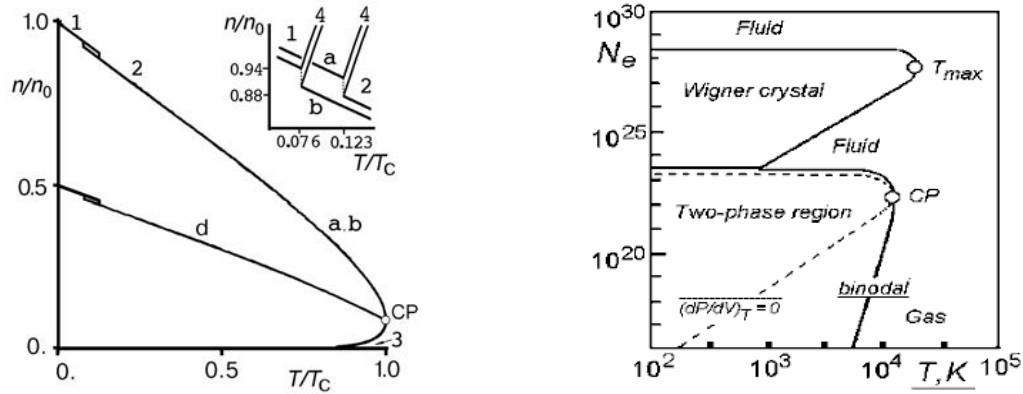

**FIGURE 2.** Density-Temperature phase diagram in two OCP($\sim$) models at low $Z \sim 1$ in the relative coordinates $n/n_0$ and $T/T_C$ where $n_0$ is the density of crystal under 'normal condition' ($P = 0$ and $T = 0$) and $T_C$ is critical temperature of fluid-gas phase transition. <u>Notations</u>: $a$ and $b$ mass-asymmetrical model of electron-ion plasma: $a$ – Single-OCP($\sim$) and $b$ – Double-OCP($\sim$) respectively; $1$ – crystal; $2$ – fluid, $3$ – gas; $4$ – melting zone; $d$ – 'diameter' of coexistence curves model $a$ and $b$; <u>Insertion</u> – structure of evaporation, sublimation and melting bounds near triple point in $a$ and $b$ models (Figure from [11]).

**FIGURE 3.** Global phase diagram in mass-asymmetrical ion-electron Double-OCP($\sim$) models of low value of charge number $Z$. Hypothetical boundary of ionic Wigner crystal and boundaries of new gas-liquid and gas-solid phase transitions are noted. <u>Notations</u>: $CP$ – critical point; $T_{max}$ – pseudocritical point of maximum melting temperature; *dashed lines* – spinodals $\{(\partial P/\partial V)_T = 0\}$ (Figure from [9,10]).

## OCP($\sim$) Phase Diagram at High Values of Charge Number

Highly anomalous structure of global phase diagram was discovered at previous study of the OCP($\sim$) at very high values of charge number $Z >> Z_2^* \approx 45$ [10,11]. In this case the melting «stripe» ($\Gamma \equiv (Ze)^2/kTa \approx 175$) crosses low-density 'slope' of two-phase coexistence domain of the new phase transition. Following features are typical for this type of phase diagram (see the variant $Z = 100$ at Figure 1 for details):
- Triple point is placed at low-density 'slope' of two-phase boundary.
- Critical point is placed at crystalline part of two-phase coexistence domain.
- Crystal-crystal coexistence of two crystalline phases, dense and expanded ones, with the same lattice occurs in OCP($\sim$) at such high values of charge number $Z$.

## Intermediate Values of Charge Number ($Z_1^* < Z < Z_2^*$)

The most remarkable anomalous phase diagram corresponds to the case when the melting line of prototype OCP(#) model ($\Gamma \sim \Gamma_{melt} \approx 175$) crosses coexistence curve of the new gas-liquid phase transition just closely to its critical point. As a result of this coincidence:
- The only phase transition exists in the model. It corresponds to the unified global crystal–fluid coexistence, i.e. continuous superposition of melting and sublimation (see Figure 4).
- There is no true critical point.
- There is no triple point.
- Coexistence curve in *P-T* (pressure-temperature) plane is a continuous, infinite curve (Figure 5).

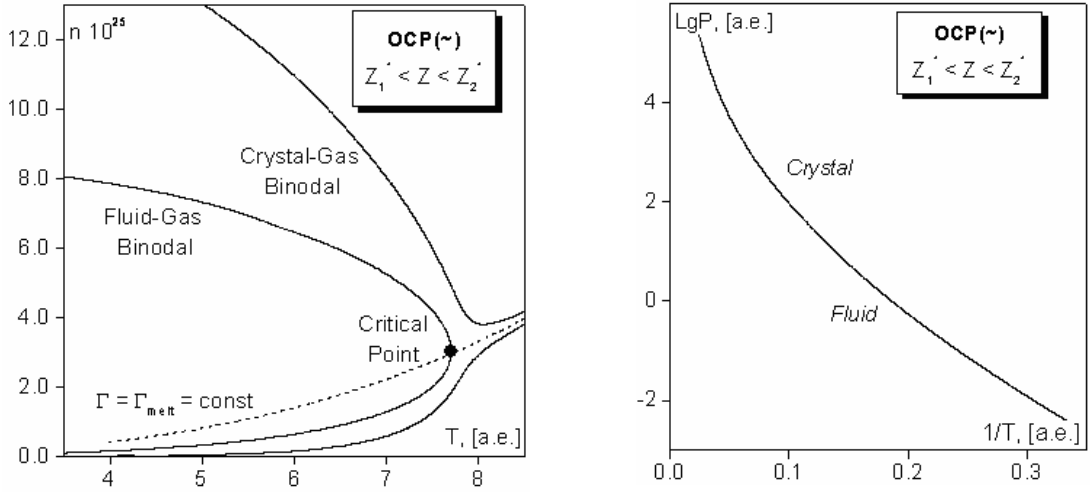

**FIGURE 4.** Anomalous type of density-temperature phase diagram in OCP(~) model at $Z_1^* < Z = 40 < Z_2^*$. (Figure from [12])

**FIGURE 5.** Anomalous type of pressure-temperature phase diagram in OCP(~) model at $Z = 40$.

## Boundary Values of Charge Number Interval ($Z = Z_1^*$ & $Z_2^*$)

Remarkable feature of phase diagram of OCP(~) model in the case $Z = Z_1^*$ or $Z = Z_2^*$ is an existence of pseudo-critical point where the well-known standard conditions are fulfilled: $(\partial P/\partial V)_T = 0$ and $(\partial^2 P/\partial^2 V)_T = 0$: at $Z = Z_1^* \approx 34.6$ – on fluid part of crystal-fluid binodal (Figure 6) and at $Z = Z_2^* \approx 45.4$ – on crystalline part of crystal-fluid binodal (Figure 7). When we use the same as in [10,11,12] analytical fits for equation of state of both subsystems, OCP(#) and background, we obtain following parameters of the both pseudo-critical points:

**TABLE 1.** Parameters of pseudo-critical point in OCP of classical point charges on the uniform and compressible background of ideal fermi-gas of electrons ($Z = Z_1^*$ or $Z_2^*$).

$\Gamma \equiv Z^2 e^2/a_i kT$ ; $r_S \equiv a_e/a_B$ ; $\theta \equiv kT/\varepsilon_F \equiv 4/(9\pi)^{1/3}(n_e \Lambda_e^3)^{2/3}$ ; $\Lambda_e^2 \equiv 2\pi\hbar^2/m_e kT$ ; $a_j^3 \equiv 4\pi n_j/3$

|  | $Z$ | $T_C$, a.u. | $(n_e)_C$, cc$^{-1}$ | $P_C$, a.u. | $\Gamma_C$ | $(r_S)_C$ | $(n_e \Lambda_e^3)_C$ | $(\theta)_C$ |
|---|---|---|---|---|---|---|---|---|
| $Z = Z_1^*$ | 34.6 | 6.38. | $2.24\ 10^{25}$ | 11.4 | 140 | 0.416 | 3.30 | 2.91 |
| $Z = Z_2^*$ | 45.4 | 9.29 | $3.96\ 10^{25}$ | 28.4 | 181 | 0.344 | 3.26 | 2.89 |

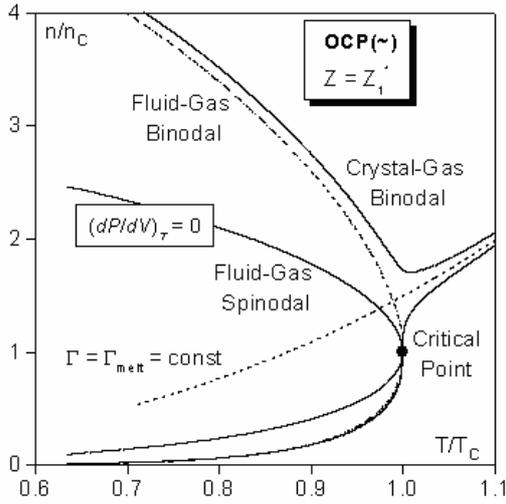 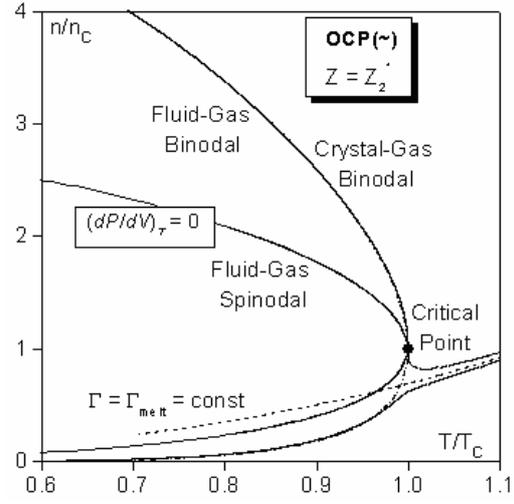

**FIGURE 6.** Anomalous type of density-temperature phase diagram in OCP(~) model at $Z = Z_1^* \approx 34.6$

**FIGURE 7.** The same as Figure 6 for the case $Z = Z_2^* \approx 45.4$ (Figures from [12]).

## *Critical exponents*

Remarkable feature of two discussed pseudo-critical points at $Z = Z_1^*$ or $Z = Z_2^*$ is the non-standard values of all critical exponents in comparison with the ordinary (Van der Waals like) critical exponents, which correspond to the case of OCP(~) with the charge number Z beyond the discussed interval $Z_1^* \div Z_2^*$. For example, at the latter case $Z < Z_1^*$ or $Z > Z_2^*$, the standard density-temperature relation is valid for the density at coexistence curve $\rho(T)$ $[\rho(T) - \rho_C] \sim |T - T_C|^{1/2}$. At the same time in the case $Z = Z_1^*$ or $Z = Z_2^*$ at pseudo-critical points the following relation may be proved $[\rho(T) - \rho_C] \sim |T - T_C|^{1/3}$ (see [12]).

## GENERAL CHARACTER OF ANOMALOUS PHASE DIAGRAMS

Anomalous phase diagram with an unique phase equilibrium crystal–fluid is not a exclusive feature of OCP(~) model. The same behavior is quite common for some systems with traditional inter-particles interaction, which combines intensive shot-range repulsion and finite in depth and spread attraction. For instance, transition from "normal" phase diagram to anomalous one has been observed in the one component system of hard spheres with additional short-range Yukawa-like attraction:

$$V(r) = \begin{cases} \infty & r < \sigma \\ -\varepsilon \dfrac{\exp[k\sigma(1 - r/\sigma)]}{r/\sigma} & r \geq \sigma \end{cases}$$

Analytical and numerical modeling of this system along with experimental results (large colloidal particles in polymer solution) show the effect qualitatively similar to one in OCP(~) model when ions charge $Z_i$ tend to the interval $[Z_1^* \div Z_2^*]$. That is gradual closing triple and critical points right up to their merging and formation a single crystal-fluid phase boundary (Figure 8(*a,b*)).

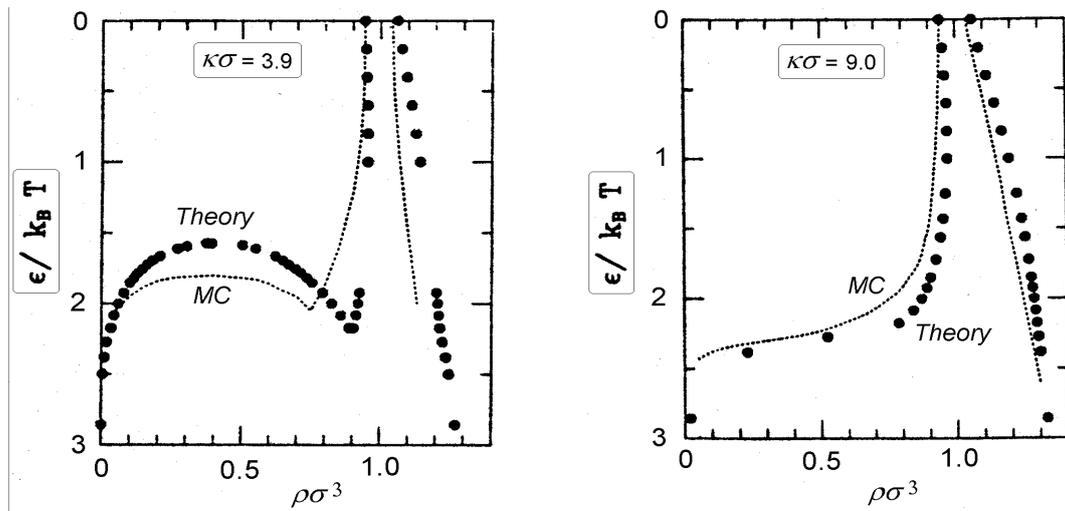

**FIGURE 8**(*a,b*). Phase diagram of hard sphere system with additional Yukawa attraction. *a* – standard type of gas-liquid-solid coexistence; *b* – anomalous phase diagram with unique crystal-fluid transition. Notations: *Solid circles* – calculation via thermodynamic perturbation theory (TPT); *dotted line* – numeric simulations results (Figure from [14]).

## CONCLUSIONS

Anomalous phase diagrams are quite common phenomena in idealized Coulomb systems with high values of charge number *Z*. Similar topology of phase diagram could be expected at some conditions at astrophysical objects. Using rather simple but modified plasma models it is possible to calculate explicitly all parameters of such anomalous phase transitions and clarify the peculiar topology of its phase diagrams.

The authors are grateful to H. DeWitt, D. Yakovlev and A. Potekhin for fruitful discussions..